# Mathematical Modelling of the Thermal Accumulation in Hot Water Solar Systems

Stanko Vl. Shtrakov, Anton Stoilov
South - West University "Neofit Rilski", Dept of Physics, Blagoevgrad, BULGARIA,
E-mail: sshtrakov@abv.bg, antonstoilov@abv.bg

**Abstract** – Mathematical modelling and defining useful recommendations for construction and regimes of exploitation for hot water solar installation with thermal stratification is the main purpose of this work. A special experimental solar module for hot water was build and equipped with sufficient measure apparatus. The main concept of investigation is to optimise the stratified regime of thermal accumulation and constructive parameters of heat exchange equipment (heat serpentine in tank). Accumulation and heat exchange processes were investigated by theoretical end experimental means. Special mathematical model was composed to simulate the energy transfer in stratified tank. Computer program was developed to solve mathematical equations for thermal accumulation and energy exchange. Extensive numerical and experimental tests were carried out. A good correspondence between theoretical and experimental data was arrived.

## 1. Introduction

Solar hot water installations became popular solar applications in Bulgaria for the last years. There are good climatic conditions in Bulgaria for all seasons' application of solar energy in domestic and public sector. For small scale solar installations, preliminary used in domestic sector, thermally stratified storage tanks for hot water are good installation scheme. In such systems the hot water remains separated from the cold water by means of buoyancy forces. Stratified storage tanks are more thermally and economically effective. Upper water layers in accumulator, where the consumption of hot water is realized, are always with higher temperature than the bottom layers. Inlet working fluid for the solar collectors is usually taken from the bottom region of accumulator (the coldest temperature) and this determines maximal energy efficiency for solar collectors.

Maintaining thermal stratification is very important. Degradation of thermal stratification in storage tanks is caused by different thermal mechanisms (Zurigat et al.,1989; Shahab A., 1999). The first mechanism is the forced convection flow through the tank. Inflow and outflow streams in the tank during the charging process and momentum-induced mixing between the incoming and resident water are a major cause of destratification. The next reason can be the natural convection caused by incoming a colder fluid from solar collectors in the top of thermal accumulator, where it is the highest temperature layer. This depends on the weather conditions (clouds) and the fluid flow rate in collector circuit.

For all seasons' applications it is necessary to use an indirect solar scheme for heat accumulation, because a special unfreezing work fluid must be used. In small solar installations this can be realized by serpentine heat exchanger, mounted in volume of water accumulator. Free convection heat exchange process is in main importance for solar energy conversion in such solar installations. The accumulator shape is also of great importance for water temperature stratification along the accumulator height. The accumulator must be vertically situated and height is the main constructive parameter assuring good and sustainable temperature stratification.

Solar installations with stratified tanks and heat exchange by included serpentine have advantages, because the stratification's degradation, caused by fluid mixing in charge phase is practically eliminated. Only in discharge process the water mixing is available, but with some constructive measures the losses in stratification can be minimized. Moreover, the place of heat exchange process in the tank can be regulated with a serpentine location. The serpentine can be situated in upper, middle or bottom part of the tank. What is the right position of the heat exchanger (serpentine) is a disputable question now. It depends on many factors and studies in this field will help designers and constructors on solar installations for hot water.

When the serpentine is mounted in the bottom part of the tank, a high energy efficiency of solar collectors can be realized. It is because the relatively low temperature in the bottom layers in the accumulator extracts maximal heat from working fluid, circulating through the collectors. Thereby the inlet temperature for solar collectors is relatively low and energy efficiency will be high. In this case however, considerable thermal stratification can't be reached, because the free convection transfers the heat upwards in all volume of accumulator.

In contrary, if the serpentine is located at the top part of the accumulator, the maximal thermal stratification can be realized. It is because the heat is accumulating in upper water layers and hot water with lower density is not transferring down to lower water layers. However, this also is not a good strategy for accumulation scheme in the solar installation, because the bottom part of the accumulator is not efficiently used (really it does not take part in solar accumulation). The inlet temperature for solar collectors will be higher and energy efficiency will be low.

Such a situation is in fact also, when the serpentine is mounted in the middle part of the accumulator, because the volume under serpentine is thermally separated from heat accumulation process.

All that indicates that the efficient way for solar accumulation with thermal stratification needs the serpentine to be divided in two or three parts. These must be located in different regions along the height of the tank – in the top of accumulator for ensuring the quick temperature rising by heat exchange with outgoing working fluid from solar collectors and in the bottom part for fully extracting the heat from working fluid.

Temperature stratification depends on the fluid flow rate in solar collectors' circuit. If the flow rate in collectors is high, a small temperature increase of the working fluid is performed and significant temperature difference between upper and down water layers in the accumulator can't be realized. Also, the small temperature difference between inlet and outlet collectors' fluid temperature increases possibilities for occurring a cold convective stream in the top of the tank, when outlet collectors' temperature is lower than water temperature in the top layer. On the other hand, low value of flow rate in collector circuit leads to higher temperature difference between inlet and outlet fluid temperature in solar collectors and possibility to accumulate water with high temperature in the top part of the accumulator. Small flow rate in solar collectors however, leads to decreasing energy efficiency of solar collectors.

Energy efficiency of thermally stratified accumulator depends more sensitively on the consumption regime of hot water in solar installation. It is important whether the hot water consumption is performing in short time (restaurant, school regime) or it is prolonging the long period of the day and night (home or hotel regime).

Investigation of commented characteristics and parameters for solar installation is very difficult, because the experiments need a long time and numerous variants of constructive and regime parameters' combinations. One useful solution for this problem would be developing a mathematical model, which has to be verified by many experiments in a wide range of variation of parameters.

Motivated by this point, the present study is intended to investigate performance of typical domestic hot water installation in different regimes of thermal accumulation and climatic conditions. In order to wide the scope of investments, a computational model and computer program for heat exchange process in accumulation tank was created. Many experimental and numerical studies have been conducted on the performance of stratified storage tank under different operating conditions and for different design characteristics. These characteristic include the parameters and situation on the heat exchange serpentine, flow rate in collector circuit, water consuming regime and other.

A variety of models and experiments to assess the efficiency of stratified tanks has been produced in many laboratories and functioned solar installations. As a result many numerical and experimental studies have been conducted on the performance of stratified tanks under different operating conditions and constructive parameters. The most of published studies analyze the direct solar installations or indirect installation with removed heat exchanger, where mass and thermal transport mechanism in accumulator and heat exchanger are separated. Solar installations with serpentine exchanger, located in water accumulator, are investigated rarely, especially with assessing temperature stratification. In such solar installations the thermal exchange process and heat accumulation perform simultaneously at the same place. This leads to the appearance of some special physical effects, which define different energy efficiency in exploitation period. Experimental and theoretical investigations, made in this work are intended to wide the knowledge for commented above problems.

**2. Experimental apparatus and measure instruments**.
The test solar system is installed on the flat roof of South-West University "N. Rilski" (SWU) – Blagoevgrad. The design and realization of the system were oriented to simulate different regimes of solar energy conversion and hot water consumption. The main target is to ensure possibility to: work in direct and indirect regime with different heat exchange area and location of serpentine unit in the tank; change angle of collectors toward by horizon and azimuth; regulate some operating parameters (fluid flow rate); measure all important parameters, influencing the system performance.

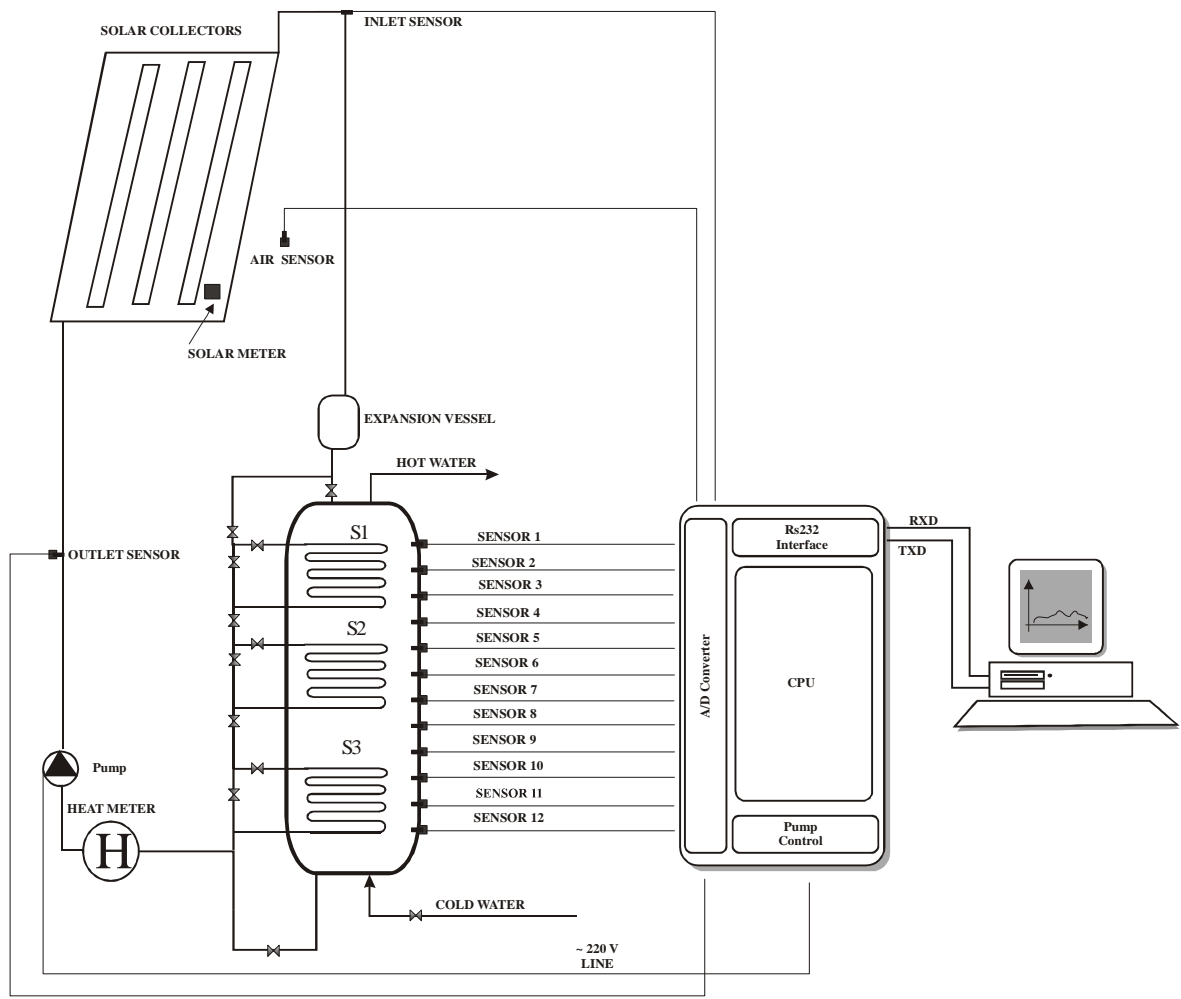

**FIG.1 Schematic diagram of the test apparatus**

The test tank is a vertical cylindrical vessel made of stainless steal material. The tank height is 1.7 m with an internal diameter of 0.35 m. The volume of the tank is 160 l, which is a typical water capacity for a family house. At the top of the tank there is outlet fitting for water consuming. Other fitting is located at the bottom of the tank and it allows injection cold water from water supply net. The tank and connection pipes are well insulated.
In the tank are built-in three copper serpentines along in the all height of the tank. Serpentines are 10 meters each in length. They can be switched on or off as a heat exchange unit by a system of valves. So the system can work with one, two or three serpentines situated in different regions of the tank. Installation can work also in direct regime, when all three serpentines are turned off.

The system is equipped with a flat solar collector 2 m² in area. The system is equipped with expansion vessel, pump, valves and other additional elements.

Monitoring system includes 12 thermo sensors assembled in the accumulation vessel, 6 thermo sensors in collector circuit, one thermo sensor for measuring the ambient air temperature. Solar meter, located near solar collectors measures solar radiation. The inflow rate, heat energy and heat power are measured by a combined heatmeter. All observed parameters are registered by an automatic monitoring system. It includes a special electronic module for converting the analog data from sensors to digital signals. Digital data from converting module is collected by the computer system. After that, the stored data can be used for detailed analysis of thermal and economical efficiency of the system and preparing the statistical calculation for long-term analysis.

Measuring module includes also a control unit, which governs the pump performance.

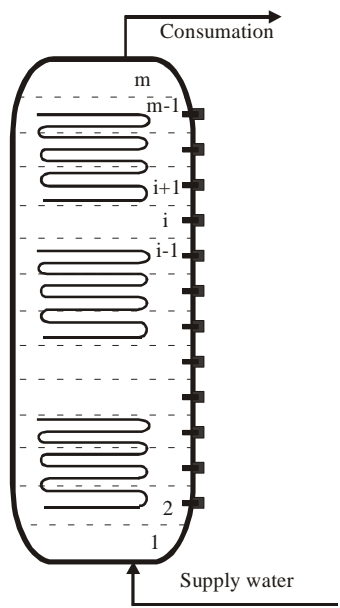

Fig.2. Schematic diagram of the tank

### 3. Theoretical modeling

Numerical modeling of thermal storage tanks is often used computational technique to investigate thermal and mass processes. Although two or three-dimensional analysis are possible to describe flow and temperature distribution in storage tank, they are not applicable to simulation calculations of the long-term performance due to insuperable difficulties of calculation algorithms. This is especially true for the transient behavior of the tank performance. Hence, one-dimensional modeling is possible alternative, because of its simplicity and sufficient accuracy of computational procedures in cases of stratified accumulators when the height of the accumulator is bigger than its diameter (for example $H/D = 3 – 5$).

A simple one-dimensional numerical model has been developed for predicting the transient behavior of the vertical temperature distribution in the tank. The model describes temperature changing in different layers of the tank by means of momentary energy balance for defined quantity of water.

The stratified accumulator, considered in this study, is divided into *m* sectors (these may be corresponding to the number of thermo sensors) with the equal volume, as depicted schematically in Fig.2. The sectors are numbered from the bottom to the top of the tank. Different sectors contain different parts of serpentine. This means that in sectors act different heat sources (heat exchange area). The intensity of heat, transferred by the serpentine to the water in the tank, decreases from up to down, because the temperature of working fluid is decreasing by heat extracting.

Heat and mass transfer processes in water accumulators for hot water solar installations are very complicated, because of mixing different mechanisms of heat and mass exchange in volume – water consumption, natural convection locally round the serpentine and globally in all the volume, heat conductivity in water, forced convection in serpentine and heat losses to the ambient. It is impossible all this processes, to be described in one mathematical system. In such complicated physical systems special Operator Splitting Method and splitting schemes have to be applied to split the general model into separated sub-problems. The splitting scheme is based on the well-known in physics superposition mechanism, which decouples heat and mass transfer phenomena and difficulties associated with non-linearity in the most of the used equations. It means that different heat and mass transfer mechanisms are modeled separately and act consequently in given time period.

*The first* heat and mass transfer mechanism is the hot water consumption. Hot water is consumed from upper sector of the tank. Consumed water quantity is compensated by injection cold water at the bottom sector. This water is assumed to mix with the water in the sector. Some quantity of water from the bottom (first - 1) sector enters the next upper sector (second - 2) and mixes with water in the sector. This process occurs in all next sectors of the tank. The temperature change in sectors by discharging process can be written by:

$$T_{i,n} = [(V_i - \nabla V)T_{i,n-1} + \nabla V T_{i-1,n-1}]/V_i, \qquad (1)$$

where $i,n$ are sector and time step number; $V$ is volume and $T$ - temperature of water. $\Delta V$ is quantity (volume) of consumed water for time step $n$.

For the bottom sector the temperature $T_{i-1,n-1}$ is the net supply water temperature $T_{net}$. Discharging process is simulated by a sequent passing the sectors from the bottom to the top. This process acts not constantly. There is a consumption graph, which represents the daily distribution of hot water.

*The Second process* in the accumulator is the thermal charging process. It is acting all the time when the solar radiation is available. The heat from solar collectors is transferred to the water in the tank by serpentine elements. This causes temperature rise of the water in the tank. Temperature rise depends on outlet temperature from solar collectors and flow rate of the working fluid. The charging process is considered as independent (Operator Splitting Method). Hence, a second passing across the sectors for the same time step is needed to determine the temperature rising. Energy balance in sectors gives the temperature change:

$$T_{i,n} = T_{i,n}' + \frac{K_{i,ser} F_{i,ser}}{\rho V_i c_p}(T_f - T_{i,n}')\nabla \tau, \qquad (2)$$

where $T_{i,n}'$ is the temperature in $i$-sector of the accumulator and $n$-time step, after the discharging process has passed; $T_f$ - average fluid temperature in $i$-serpentine element; $\Delta \tau$ - time step interval for charging process; $K_{i,serp}$ and $F_{i,serp}$ - heat transfer coefficient and heat exchange area of serpentine element for $i$-sector; $\rho$ and $c_p$ - density and specific heat capacity of water in tank.

Heat transfer coefficient $K_{i,serp}$ is determined by convection transfer coefficients for two sides of the serpentine. External surface of the serpentine transfers heat to the water in accumulator by natural convection. Convective transfer coefficient $h_{free}$ is received from:

$$h_{free} = \frac{Nu_{free} \lambda_s}{d_o}, [\text{W/m}^2 \text{ K}] \qquad (3)$$

where $Nu$ is Nuselt number, $\lambda_s$ - conductivity coefficient of the water in accumulator, and $d_o$ - external diameter of serpentine pipe.

For the Nuselt number determination is used criteria equation:

$$Nu_{free} = 0.394 Gr^{0.2} Pr^{0.25} \qquad (4)$$

where $Pr$ is Prandtl number, $Gr = \dfrac{\beta \cdot g \cdot d_o^3 \cdot (T_s - T_{acum})}{\nu^2}$ - Grashoff number, $\beta$ - heat expanding coefficient, $g$ - earth gravity, $T_s$ - temperature of serpentine wall, $T_{acum}$ - temperature of water in accumulator, $\nu$ - fluid viscosity.

Working fluid in the serpentine transfers heat to the serpentine wall by forced convection with convective transfer coefficient $h_f$:

$$h_f = \frac{Nu_f \lambda_f}{d_i}, [\text{W/m}^2 \text{ K}], \qquad (5)$$

where $Nu_f = 2.1 \cdot 10^{-2} \cdot Re_f^{0.8} \cdot Pr^{0.43} \cdot \left(\dfrac{Pr_f}{Pr_s}\right)^{0.25}$ is the Nuselt number and $Re_f$ - the Reinolds number for the working fluid.

Indexes in the last part of $Nu$ equation refer to the Prandtl numbers calculated for temperature of fluid and temperature of serpentine wall, respectively.

Overall heat transfer coefficient $K_{i,serp}$ for serpentine element includes convective coefficients $h_f$ and $h_{free}$ and conductive transfer parameters for serpentine wall:

$$K_{i,serp} = \frac{1}{\frac{1}{h_f} + \frac{\delta_s}{\lambda_s} + \frac{1}{h_{free}}}, \; [\text{W/m}^2 \text{ K}] \qquad (6)$$

Convective coefficients depend on fluid temperatures, which are unknown values in the beginning of calculations. Known parameters for calculation start are the inlet fluid temperature for serpentine (outlet collector temperature) and water temperature in the accumulator (temperature distribution in accumulator). Initial temperature distribution in the accumulator must be adopted in the beginning of the calculations (initial conditions). This predestines the calculation consequence - from the top to the bottom of the accumulator because the inlet of the serpentine is in the top region of the accumulator. Calculation begins for the top accumulator sector with the known fluid temperature in entrance of serpentine element.

Iteration procedure for transfer coefficient $K_{i,serp}$ is adopted in this work. In the first step fluid parameters are determined with approximate calculation of outlet temperature of working fluid for serpentine element:

$$T_{i,out} = T_{i,in} - \varepsilon \cdot (T_{i,in} - T_{i,ac}) \qquad (7)$$

where $\varepsilon$ is exchange efficiency coefficient for serpentine element and $T_{i,in}$ is known inlet temperature for the serpentine element.

Exchange efficiency coefficient is not available on this stage of calculation; therefore an initial value can be adopted. Numerical experiments we have carried out show that values in range 0.5 – 0.7 give a good initial approximation for iteration process. Average fluid temperature in serpentine element now can be calculated:

$$T_{av,i} = \frac{T_{i,in} + T_{i,out}}{2}, \qquad (8)$$

This temperature is used to calculate fluid parameters of working fluid in serpentine. For Grashoff number it is necessary to know serpentine wall temperature $T_s$. This temperature depends on coefficients $h_f$ and $h_{free}$ and it is near to the working fluid temperature because the $h_f$ coefficient is greater than $h_{free}$. In this work, as initial value for $T_s$ has been used the next approximation:

$$T_s = T_{i,av} + 0.2 \cdot (T_{i,av} - T_{i,ac}) \qquad (10)$$

Now, all needed data for calculation of heat transfer coefficient $K_{i,serp}$ is presented and the first iteration step can be performed. Heat flux between serpentine and water in accumulator sector and the new temperature of water in accumulator can be calculated by:

$$q = F_i \cdot K_{i,serp} (T_{f,a} - T_{ac,i}) \qquad (11)$$

$$T_{i,ac} = T_{i,ac}' + \frac{q}{\rho V_i c_p} \nabla \tau \qquad (12)$$

where $T_{i,ac}'$ is initial water temperature in sector.

Now the outlet temperature of working fluid can be calculated again by using the actual heat transfer flux:

$$T_{i,out} = T_{i,in} - \frac{q}{\rho V_i c_p} \nabla \tau \qquad (13)$$

This is not the last calculation for outlet temperature for serpentine element, because the initial parameters in calculation procedure were only approximate. It can be used further for determining the average temperature of working fluid in serpentine element – equation (8).

Calculations (8) - (10) with considering (2) - (6) can be repeated with temperature $T_f$ instead $T_{av,i}$ (eqn. 7) and also water temperature in accumulator can be corrected: $T_{i,ac} = (T_{i,ac}+T'_{i,ac})/2$. New heat flux and outlet temperatures will be received and calculations can be continued until difference between temperatures in two consecutive calculations become very small.

Inlet fluid temperature $T_{i,in}$ in *i*-sector is known – this is the outlet temperature from the previous (upper) serpentine element (sector *i+1*). It stays constant in the iteration process. For the top sector inlet fluid temperature in serpentine element is determined by solar collector's performance. The outlet temperature $T_{i,out}$ depends on transferred heat energy in the sector and it is determined by equation (13) in last iteration step.

Mathematical model for solar collectors is well-established matter and detailed information can be found in solar energy publications [1]. Outlet temperature of working fluid for solar collectors is defined by next equation:

$$T_{sol,out} = \frac{F_R}{m \cdot c_p}[q_s(\tau \cdot \alpha)_e - U_L(T_{sol,in} - T_a)] \qquad (14)$$

where $F_R$ is heat removal factor, *m* – mass flow rate of working fluid, *(τα)$_e$* – effective transmittance absorbing coefficient for optical part of solar collectors, $q_s$ – solar radiation flux for tilted surface [W/m$^2$], $U_L$ – overall collector heat loss coefficient [W/m$^2$ K], $T_a$ – ambient temperature.

Inlet temperature for solar collectors $T_{col,in}$ is formed by the outlet temperature from the bottom serpentine element. This temperature is not available in beginning of the charging calculation process. Dependence between inlet temperature of working fluid for serpentine and inlet temperature for solar collectors (outlet temperature for serpentine) presumes a new iteration calculation to be organized. This can be performed by repeating all serpentine calculations (6) – (14) with a new inlet temperature for solar collectors $T_{i,out}$, received from calculation of last serpentine element. For starting this procedure there is need to adopt an initial value for outlet temperature of working fluid from the serpentine. Because the serpentine is taken to be with enough exchange area the initial value of outlet temperature can be taken equal to the water temperature in the bottom sector of the accumulator or little above this temperature (1-2 $^o$C). Our experience shows that this iteration is very fast. Only two or three passes are needed for receiving proper temperatures in the system.

*Next process* which must be analyzed for temperature distribution in solar accumulator is thermal conductivity in water. This process determines the microscopic heat transfer between substance layers, if their temperatures are different. As it is known, unsteady thermal conductivity can be described by general conductivity equation. For solar water accumulators with thermal stratification the accumulator height is usually bigger than the accumulator diameter and temperature distribution is in main importance among the height. This means that one dimensional heat conductivity equation (among the height) is acceptable approximation for describing thermal conductivity in the accumulator:

$$\frac{\partial^2 T}{\partial z^2} = \frac{1}{a}\frac{\partial T}{\partial t}, \qquad (15)$$

where $a = \frac{\lambda}{c_p \rho}$ is temperature conductivity number; *z* – coordinate in height of accumulator; *t* – time. *T* is temperature in accumulator after discharging and charging process has been performed for given time step.

This equation can be solved numerically by using calculation mesh formed by dividing the accumulator in sectors (fig.1). Every sector forms a node in calculation mesh and its temperature, received from former calculations (charging and discharging processes) is the initial temperature value for numerical calculation. Numerical procedure for solving equation (15) is not presented here because the size of description exceeds limits of this work. Detailed information for numerical procedures can be received from [5,6] and from authors.

Thermal conductivity calculations leads to changes in temperature distribution in sectors in any extend, but total heat content is not changing, if it is considering the insulated accumulator. This process has small influence for thermal and mass transfer processes in active charge/discharge period, but it has some role for temperature equalization in night periods.

*Heat losses* from external walls of vessel to the ambient are calculating separately by using the standard calculation procedure:

$$Q_i = U_{loss} \cdot F_{ext} \cdot \Delta T_i \qquad (16)$$

where $U_{loss}$ is heat loss coefficient (include heat insulation of accumulator); $F_{ext}$ – external tank area for every sector, $\Delta T_i$ – temperature difference between temperature of water in *i*-sector and ambient temperature.

Heat losses results in decreasing the temperature in sectors:

$$T_{i,acum} = T_{i,acum} - \frac{Q_i}{\rho \cdot V_i c_p} \nabla \tau, \qquad (17)$$

Very important process for solar stratification in water accumulators is *natural convection* in volume. It is in action when the temperature in lower water layers is higher than the temperature in upper layers. Such situation is available in cases when the serpentine is located in the bottom or the middle part of the accumulator and heat is raised upward by free convection. Another situation with free convection is occurring when a colder working fluid enters the serpentine in the top of the accumulator. Because the temperature in upper layers is highest (in stratified accumulator) the colder fluid in serpentine will cause the downgrade cold stream in accumulator. This is a free convection process, destroying temperature stratification in the tank. Such situation happens in cases, when the solar radiation is insufficient (cloud, big incidence angle) to rise temperature above temperature in the top layers in the accumulator.

Free convection in vertical cylinder filled by water can be described by next criteria equation [3]:

$$Nu_{avr} = 0.0556 \cdot (Gr \cdot \Pr)_H^{1/3} \qquad (18)$$

**H/d** = 0.22 ÷ 1.75 ; **Pr** = 5.7 ÷ 6.65 ; **$Ra_H$** = **Gr.Pr** = $10^8 \div 2.26 \cdot 10^{11}$.

This equation describes a free convection heat transfer in cylindrical tank with insulated walls, where heat is entered in the bottom of the cylinder.

For laminar and turbulent free convection heat transfer can be used another equation [4]:

$$Nu = 0.11 \cdot (Gr \cdot \Pr)^{0.33} + (Gr \cdot \Pr)^{0.1} \qquad (19)$$

for **Gr.Pr** = $10^{-7} \div 10^{12}$

For stratified solar accumulators criteria equations mentioned above can be used for assessing the upward heat transfer between adjacent layers, having different temperatures. This acts in the case of location the serpentine in the bottom part of the tank. By means of criteria equation it can be determined free convective coefficient for heat flux definition:

$$Q = \alpha_{free} \cdot F_{sect} \cdot \Delta T \qquad (20)$$

where $\alpha_{free} = \dfrac{Nu \cdot \lambda}{h_{layer}}$ $\alpha_{free}$ is convective coefficient, $F_{sect}$ – cross-section area of accumulator and $\Delta T$ – temperature difference in water layers, $h_{layer}$ – height of water sector.

Heat flux determines temperature changing in two neighbouring layers. Lower layer's temperature will decrease and upper layer's temperature will increase with the same value. By consecutive calculations from the bottom to the top of tank it is possible to determine temperature distribution change in the tank with considering natural convection. This technique is approximate because the natural convection is a global process for all volume of the accumulator. However, if we take into account, that the temperature distribution in accumulator is not regular (because of charging and discharging processes and heat conductivity), often there are separated and limited natural convection regions in the tank. On the other hand, discharging process (water consumption) is accompanied with mass transfer from the bottom to the top of the accumulator. In this process it is performed a heat transfer, which realizes a global heat exchange, which in many cases overcomes the global results of natural convection. This thesis is corroborated by results from many experiments we have made in last two years.

Downgrade stream formed by means of cold water, forming in upper layers when the cold working fluid incomes the accumulator from solar collectors (when the serpentine is located in the top part of vessel) is rare event. There has not be discovered a proper model for such physical phenomena in literature. Because temperature difference in adjacent layers is small a big downgrade streams can not be formed. As the first approximation, we have used the above mentioned natural convection criteria equations (18) and (19).

## 5. NUMERICAL EXAMPLES

Special simulation algorithm has been developed to bind the collector and accumulator models in a working unit. It takes into account the heat losses in installation's pipes. A computer program is created to manage the theoretical calculations.

To verify the applicability of the above proposed technique, a large number of numerical examples have been carried. The splitting scheme of mathematical model allows making verification of different physical processes of thermal accumulation. For example, if the serpentine is located in the bottom part of the vessel and there is not hot water consumption the natural convection in accumulator can be investigated separately from the other processes. It is the main process, which transfers heat from down layers upward and determines temperature distribution in accumulator in this case. Other processes play secondary role in integral performance of solar installation. Figure3 shows the temperature distribution in two layers (top and bottom), received by experiments and calculations for summer period (august) and without water consumption. Results show, that theoretical model (natural convection sub-problem) describe correctly tendencies and specificity of physical process. Many numerical experiments and real tests, we have carried out, confirms possibility for using the criteria equation (19) for calculation the temperature change in vessel, when heating is produced in bottom of accumulator. A good theoretical results have been received for cases of very intensive thermal accumulation (high level of solar radiation and ambient temperature) by next criteria equation [3]: $Nu_{avr} = 0.59 \cdot (Gr \cdot \Pr)_H^{1/3}$.

More precisely this process could be treated by two-dimensional model of Navier-Stocks fluid dynamic equations.

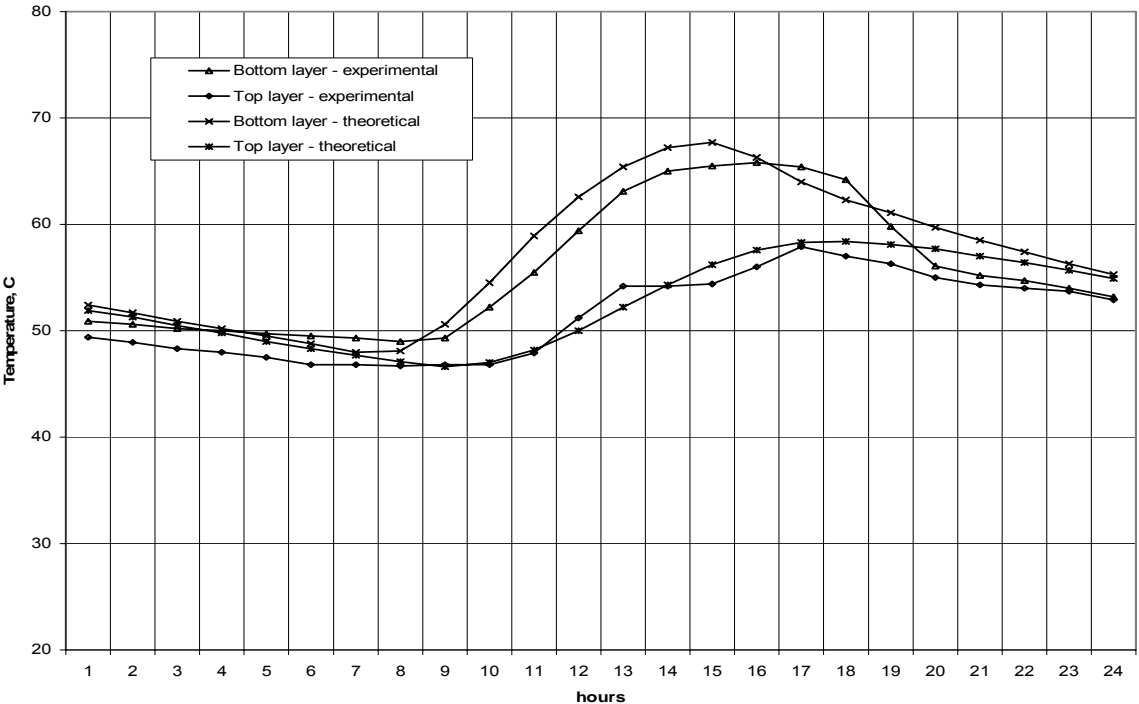

Fig. 3 Temperature distribution - serpentine in bottom

Results presented in fig.3 shows also the accuracy of other sub-problem in mathematical treatment. In time period between 22 and 7 o'clock there are not other thermal processes but heat loses to the ambient. Curves tilt in this period is the parameter, which is defined by thermal insulation of the accumulator. Presented results illustrate that heat loss coefficient $U_{loss}$ for the theoretical calculations is taken with higher value than real value (curves tilt for theoretical calculations is bigger then tilt of experimental curves).

In real exploitation of solar installations inaccuracy in description of natural convection is smaller, because this process is combined with water consumption, which performs mass transfer from down layers to the top of vessel. Water consumption is modelled almost exactly and natural convection in periods of consumption is only auxiliary process.

Thermal conductivity in water can be tested by experiments, implemented for serpentine location in top part of accumulator. In this case heat from serpentine is transferred to down layers only by thermal conductivity. This process is basic also in night hours, when hot water consumption and heat charging is missing. Experiments show that heat conductivity is well modelled by standard mathematical equation (15).

In Fig.4 is shown comparisons between experimental and theoretical results for temperature distribution in stratified accumulator in real exploitation of solar installation. Climatic data are measured values for solar radiation and ambient temperature for Blagoevgrad region, Bulgaria during the autumn (October). Results in figure correspond to the case of serpentine location in the bottom part of the vessel. Besides a good coincidence between experimental and theoretical results, here can be analyzed the role of consumption process. Despite the serpentine is located in bottom part of the tank, there is time interval with distinct thermal stratification (from 22 to 9 o'clock). This temperature redistribution (heat is realized in the bottom part of vessel, but top layers has a higher temperature) is realized by means of hot water consumption. When a hot water is consumed from the top of the tank, water from down layers with high temperature is moved above. Only in solar active period (9 – 19 o'clock) the accumulator is with thermally unstratified temperature distribution.

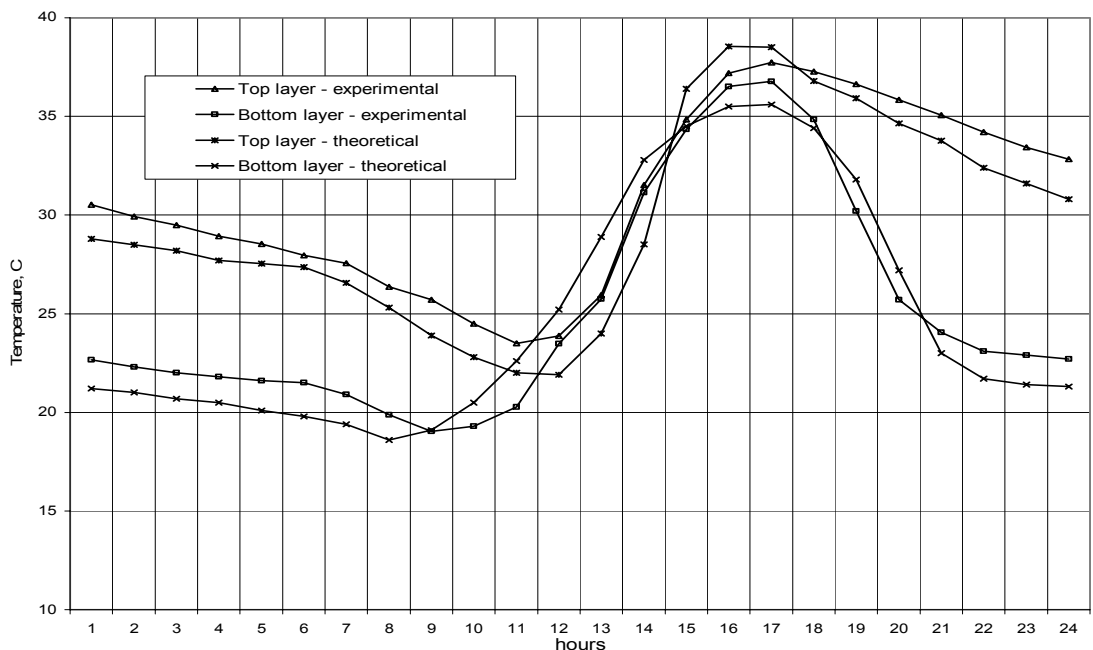

Fig. 4 Temperature distribution - serpentine in bottom part

The next figure (fig.5) shows a comparison between theoretical and experimental data for thermal accumulator with two serpentines (in the top and the bottom part of vessel). Climatic data (solar radiation and ambient temperature) are for September. Water consumption is 250 l/day with standard daily distribution [1]. Temperature stratification in this case is available almost for all time and energy efficiency is higher.

The set of daily distribution of solar radiation and ambient temperature is measured in 10 minutes period and such a period has been used for simulation calculations by theoretical model. Five days period of experiments and simulation calculations for every variant was applied to exclude influence of initial conditions and to assure stable results for comparision.

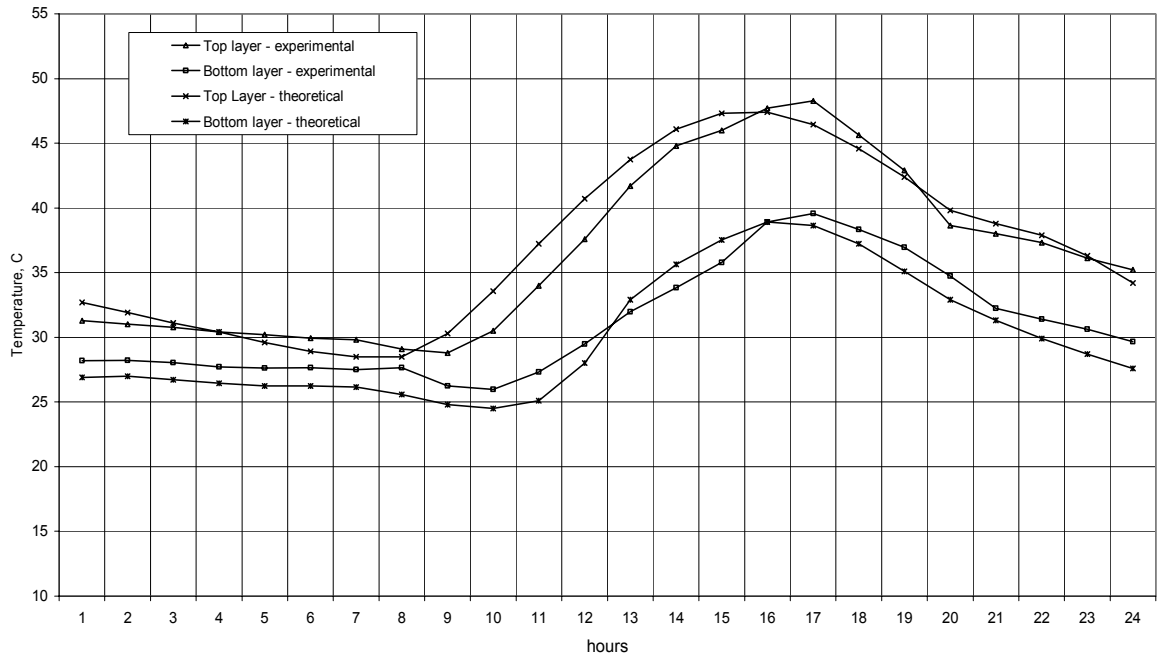
Fig.5 Temperature distribution - serpentine in bottom and top part

## 6. CONCLUSIONS

In this paper, has been presented a mathematical model for stratified accumulation in hot water solar installations with serpentine exchanger. A special Operator Splitting Method has been proposed to simplify the complicated mathematical model. A new solution procedure with several iterations schemes, suitable for the presented mathematical model was suggested.

Computer program for simulation calculations was created, and with extensive numerical experiments, the applicability of presented model was verified. Different mathematical sub-problems were validated by suitable working regimes in experimental units. A good correspondence between theoretical and experimental result have been achieved.

Mathematical model will help us to wide research area for hot water solar installation and define a useful recommendations for constructive and regime parameters of installations. This mathematical scheme is a stable base for expanding theoretical investigation by researching a two-dimensional model of stratified accumulator, working in a simulation model of solar installation. It will increase accuracy in description the natural convection in accumulator.

The results from presented mathematical model would help researches in field of stratified accumulation increase their knowledge and experience for thermal and mass transfer processes. Designers of solar installations for hot water can use results for this model to select optimal constructive parameters of the installations.

**References**
1. J.A.Duffie and W.A. Beckman, *Solar engineering of thermal processes*, Wiley Interscience, New York, 1980.
2. G.F.Csordas, A.P. Brunger, K.G.T.Hollands and M.F. Lightstone, *Plume entraintment effects in solar domestic hot water systems employing wariable-flow-rate control strategies*, Solar Energy 49 (6), 497-505 (1992).
3. Goldstein R.J. Heat transfer by thermal convection at high Rayleigh numbers, IJHMT, 1980, v. 23, No 5, p. 738-74
4. H. Hausen, *Warmeubertragung im gegenstrom, gleichstrom und kreuzstrom*, Springler-Verlag, Berlin Heidelberg New York, 1976
5. 5. Cranc J., P.Nicolson. A practical method for numerical evaluation of solution of partial differential equations of heat-conduction type. Proc.Cambridge Philos.Sos.,1974, N 43

6. Richtmayer R.D.,K.W. Morton, Difference method for initial-value problems. Interscience Publishers, 1977.